\documentclass[journal]{IEEEtran}
\ifCLASSINFOpdf
\else
\fi
\usepackage{graphicx}   
\usepackage{framed}     
\usepackage{subfig}     
\usepackage{soul}
\usepackage{xcolor}
\usepackage[utf8]{inputenc}
\usepackage{enumitem}
\usepackage{subfiles}
\usepackage{tabularray}
\usepackage{url}
\usepackage{tabularx} 

\begin{document}

\title{Establishing Minimum Elements for Effective Vulnerability Management in AI Software}
\author{
\IEEEauthorblockN{Mohamad Fazelnia, Sara Moshtari, Mehdi Mirakhorli}\\ University of Hawaii at Manoa, HI, USA\\ \{mfazel, saram23, mehdi23\}@hawaii.edu}

\maketitle

\begin{abstract}
In the rapidly evolving field of artificial intelligence (AI), the identification, documentation, and mitigation of vulnerabilities are paramount to ensuring robust and secure systems. This paper discusses minimum elements for AI vulnerability management and the establishment of an Artificial Intelligence Vulnerability Database (AIVD) supported by an AI Bill of Materials (AIBOM) for structured model documentation. It presents standardized formats and protocols for disclosing, analyzing, cataloging, and documenting AI vulnerabilities. It discusses how such an AI incident database must extend beyond the traditional scope of vulnerabilities by focusing on the unique aspects of AI systems. Additionally, this paper highlights challenges and gaps in AI Vulnerability Management, including the need for new severity scores, weakness enumeration systems, and comprehensive mitigation strategies specifically designed to address the multifaceted nature of AI vulnerabilities.

\end{abstract}

\begin{IEEEkeywords}
AI Incidents, AIVD, AI Vulnerability Management, AI Bill of Materials, AIBOM
\end{IEEEkeywords}

\section{Introduction}

As artificial intelligence (AI) technologies permeate various sectors, including healthcare \cite{stephan2002multicenter}, finance \cite{board2017artificial}, transportation \cite{lorsakul2007traffic}, industry\cite{cockburn1996archon,liang2021deep}, and security \cite{wen2017image, radanliev2022super, li2018vuldeepecker,chakraborty2021deep}, their transformative impact is undeniable. 
However, this rapid integration introduces complex vulnerabilities that may compromise the systems it is intended to enhance. 
With AI becoming a foundational component across various sectors, there is an urgent need to systematically identify, evaluate, and mitigate AI-specific vulnerabilities.
Therefore, Vulnerability Management in AI systems is a critical process that encompasses the identification, assessment, public disclosure, and remediation of security vulnerabilities within the systems and components. 
This practice is essential due to the unique and complex nature of AI software (software that contains AI components), which often includes data-driven decision-making processes that are opaque and challenging to secure.

The establishment of a robust vulnerability management framework for AI systems underscores the need for centralized information sharing, where AI vulnerabilities can be reported, cataloged, and communicated with all stakeholders. 
This centralization ensures that information regarding new and emerging threats is quickly and widely disseminated, allowing for more rapid responses across the industry. 
This demands a global AI vulnerability database to serve as a central repository for information on identified vulnerabilities specific to AI systems, providing crucial support for organizations in protecting their AI technologies against security threats. 
While the National Vulnerability Database (NVD) \cite{NVD}, along with frameworks like Common Vulnerabilities and Exposures (CVE) \cite{CVE} and Common Weakness Enumeration (CWE) \cite{CWE}, provides a well-established infrastructure for documenting traditional software vulnerabilities, these systems are designed for static, code-based flaws and lack support for the dynamic, data-driven nature of AI. They do not account for vulnerabilities emerging from model behavior, training data quality, or evolving algorithmic interactions.

AI vulnerabilities differ significantly from traditional software vulnerabilities due to the adaptive and often data-driven nature of AI systems. Unlike conventional software, where vulnerabilities typically arise from coding errors or system misconfigurations, AI vulnerabilities often stem from training data, model architecture, or learning algorithms. For instance, adversarial attacks \cite{43405}, where malicious inputs are crafted to manipulate model outputs, highlight threat vectors unique to AI. Such vulnerabilities can lead AI systems to behave unpredictably or make decisions that benefit attackers. Identifying and mitigating these issues requires analyzing not only the code but also the behavior and data that shape the model, making the process more complex and less deterministic than in traditional systems.

Given these fundamental differences, there is a clear need for a dedicated repository that focuses specifically on AI vulnerabilities. The proposed Artificial Intelligence Vulnerability Database (AIVD) is designed to fulfill this role, functioning similarly to the NVD but tailored to capture the unique characteristics of AI systems. In this paper, we outline the \textbf{minimum elements} (MEs) required for effective AI vulnerability management and propose a framework for establishing AIVD. We introduce standardized formats and protocols for reporting, analyzing, cataloging, and measuring AI vulnerabilities.

Unlike existing databases, AIVD must account for AI-specific issues such as biases in training data and flaws in algorithmic design. We also identify critical gaps in current practices, including the need for new severity scoring mechanisms, weakness classification systems, and mitigation strategies tailored to the evolving behavior of AI systems. These systems, driven by data and algorithmic complexity, require adaptive vulnerability management approaches to account for their dynamic, non-deterministic nature. Section~\ref{sec:moving-forward} further expands on these challenges and outlines the path forward.

The main contributions of this paper are five-fold:
\begin{enumerate}
    \item AI-specific Common Weakness Enumeration (AI-CWE) Definition: We introduce a structured weakness enumeration system tailored to AI systems, distinct from traditional CWE by focusing on AI-specific failure modes.
    \item MEs for AI Vulnerabilities: We define a novel set of MEs necessary for describing AI vulnerabilities across model, data, and deployment layers—supporting consistent reporting and assessment.
    \item AI Bill of Materials (AIBoM) Framework: We outline AIBOM, an AI-specific extension to SBOM, capturing not only system components and dependencies but also ethical, environmental, and usage boundaries specific to AI workflows.
    \item AI-Aware Severity Scoring: We analyze the gaps in CVSS and propose two directions—dynamic scoring and AI-specific impact metrics—for developing a more suitable scoring system for AI vulnerabilities.
    \item Gap Analysis and Future Challenges in AI Vulnerability Management: We identify and formalize the critical challenges that distinguish AI vulnerability management from traditional software security practices. 
\end{enumerate}

\begin{table*}[ht]
\centering
\caption{Details of AI Vulnerability Assessment}
\begin{tabular}{|c|p{1.6cm}|p{15cm}|}
\hline
\textbf{Item} & \textbf{Field} & \textbf{Description} \\ \hline
1 & \textbf{AI-CVE ID} & This is a unique identifier assigned to each recorded vulnerability within the AIVD fortracking and referencing. \\ \hline
2 & \textbf{AI Model Details} & Specific details of the AI model or system affected (e.g., model type, version). \\ \hline
3 & \textbf{Weakness Type} & Identifies the category or mechanism of the AI system vulnerability. Similar to CWE in NVD database, each vulnerability will be assigned with an AI-CWE. \\ \hline
4 & \textbf{Root Cause} & Explains factors contributing to the vulnerability, such as inherent model flaws, insufficient robustness safeguards, or data issues.\\ \hline
5 & \textbf{Impact} & How the vulnerability is affecting the product. \\ \hline
6 & \textbf{Severity Scores} & Quantifies the impact of AI vulnerabilities, helping prioritize and guide effective mitigation.\\ \hline
7 & \textbf{Affected Software \& Products} & Lists applications affected by the vulnerability to guide targeted remediation and risk management.\\ \hline
8 & \textbf{Exploitability} & Describes how the vulnerability can be exploited, including attack complexity, required privileges, and user actions. \\ \hline
9 & \textbf{Description} & A thorough explanation of the vulnerability, or directst to an external source for a more comprehensive description of the vulnerability. \\ \hline
10 & \textbf{Mitigation} & Outlines specific steps and strategies to effectively address and neutralize a vulnerability within an AI system, taking into account the unique characteristics and behaviors of AI technologies. It includes detailed recommendations for patching flaws, strengthening security protocols, and updating AI models to resist attacks. \\ \hline
11 & \textbf{References} & Relevant documents, papers, and reports of the vulnerability. \\ \hline
12 & \textbf{Report Date} & The date on which the vulnerability was first identified or reported. \\ \hline
13 & \textbf{Reported By} & The individual or organization that first reported the vulnerability. \\ \hline
14 & \textbf{Vendor} & Identifies the developer or the supplier of the AI system or component which is affected by the vulnerability. \\ \hline
15 & \textbf{Status} & Current status of the vulnerability (e.g., unresolved, patched). \\ \hline
\end{tabular}
\label{tab:ai_vulnerability}
\end{table*}
\section{Minimum Elements for Defining AI Vulnerabilities}
This section describes the Minimum Elements (MEs) essential for capturing and describing AI vulnerabilities effectively. Each ME represents a critical piece of information that contributes to the comprehensive profiling and understanding of vulnerabilities within AI systems. These elements facilitate structured data collection, enabling systematic tracking, analysis, prioritizing, and mitigation of vulnerabilities. These elements equip stakeholders with the necessary details to address vulnerabilities from identification to resolution.  
Each ME is strategically defined to encapsulate crucial information regarding the associated vulnerability. This approach not only standardizes vulnerability reporting across various platforms and industries, but also ensures that all stakeholders, from developers and security professionals to regulatory bodies, have access to the same comprehensive data set. Structured data collection enabled by these elements allows organizations to systematically track vulnerabilities from initial identification to final resolution. This process involves analyzing vulnerabilities for impact and exploitability, helping prioritize them for mitigation.

Furthermore, these elements serve as the backbone for creating databases and registries that stakeholders can refer to to ensure consistency and accuracy in vulnerability management. By defining these minimum required data points, the AI vulnerability disclosure process becomes more efficient, transparent, and scalable. This accommodates the dynamic nature of AI technologies and their complex ecosystems. The standardized approach streamlines mitigation and improves AI security by providing clear actionable information for rapid threat response.These minimum elements are described in Table~\ref{tab:ai_vulnerability}. Some of these minimum elements require further advancements in foundational knowledge and the creation of new frameworks and catalogs to describe AI vulnerabilities.

\subsection{Weakness Type} 

Despite their sophistication, AI systems are uniquely vulnerable to weaknesses rarely seen in traditional software. 
These vulnerabilities arise from core characteristics of AI, including its reliance on large datasets and complex, often opaque decision-making processes.
While resources like CWE catalog traditional software vulnerabilities, there is no equivalent framework for AI. 
Moreover, CWE cannot fully capture AI-specific weaknesses introduced by data dependence and algorithmic complexity. 
Because AI performance hinges on data quality and volume, any corruption or bias can severely affect outcomes. 
The black-box nature of many AI models further complicates vulnerability detection and understanding, hindering the development of robust systems.


To address AI-specific vulnerabilities, we propose an AI-CWE—a catalog of weaknesses enabling a structured approach to their identification. 
The AI-CWE builds on the AI/ML ATT\&CK framework \cite{fazelnia2022supporting}, linking each weakness to gaps in mitigation strategies. 
We define four main classes of weaknesses as follows:

\begin{enumerate}
    \item Insufficient validation mechanisms that allow malicious samples to bypass security checks and infiltrate the system.
    \item Data handling processes lack robust filtering and normalization capabilities, leading to susceptibility to noise and perturbations that compromise data integrity.
    \item The learning algorithms are vulnerable to perturbations due to inadequate resilience and adaptability in handling crafted inputs designed to mislead or corrupt the learning process.
    \item Privacy safeguards are deficient or absent, exposing the data to unauthorized access and manipulation due to inadequate encryption, anonymization, or access control configurations.
\end{enumerate}

Each weakness is categorized to support detailed analysis, root-cause tracing, and targeted mitigation. Table \ref{tab:CVE1} illustrates sample AI-CWE entries and their key elements.

\begin{table}[ht]
\centering
\caption{Example of a potential AI-CWE entry (inadequate input filtering)}
\begin{tabular}{|l|p{5cm}|}
\hline
\textbf{AI-CWE-ID} & AI-CWE-100 \\ \hline
\textbf{AI-CWE-Name} & Inadequate Input Filtering \\ \hline
\textbf{Description} & The data handling processes lack robust filtering and normalization capabilities, leading to susceptibility to noise and perturbations which compromise data integrity. \\ \hline
\textbf{Examples} & A malicious actor can add carefully designed perturbation to the input data to mislead the AI system\\ \hline
\textbf{Severity} & High to Critical \\ \hline
\textbf{Common Consequence} & Affects the integrity of the model's output without manipulating the model itself. \\ \hline
\textbf{Relationships} & None \\ \hline
\textbf{Modes of Introduction} & Inference \\ \hline
\textbf{Potential Mitigation} & Adversarial Detection \& Input Reconstruction \\ \hline
\textbf{References} & \url{https://ieeexplore.ieee.org/document/9010939} \cite{9010939}, \url{https://openreview.net/pdf?id=SyJ7ClWCb} \cite{guo2018countering}\\ \hline
\end{tabular}
\label{tab:CVE1}
\end{table}


\subsection{Impact} 

Describes the repercussions of a vulnerability within the AI system, explaining how it affects functionality, security, and reliability. 
It covers both direct and indirect effects, including performance degradation, exploitation risks, and broader impacts on users and services.

\subsection{Severity Scores} 
The Common Vulnerability Scoring System (CVSS) is a widely used standard for assessing software vulnerability severity. However, in AI systems, unique characteristics such as learning behavior and data dependence reveal gaps in CVSS and similar frameworks. Given the distinctive nature of AI vulnerabilities, a tailored severity assessment approach is needed. Below, we outline these gaps, discuss measurement methods, and propose future directions.

\subsubsection{CVSS Base Metrics and AI}
\begin{itemize}
    \item \textbf{Exploitability Metrics} (Attack Vector, Attack Complexity, Privileges Required, User Interaction): Define how a vulnerability can be exploited. AI systems add complexities not captured here, such as subtle data manipulation influencing model behavior.
    \item \textbf{Impact Metrics} (Confidentiality, Integrity, Availability): While covering direct impacts, AI systems also face indirect effects—like decision errors or data poisoning—that degrade operational integrity beyond data loss or downtime.

\end{itemize}

\subsubsection{Environmental Metrics and AI}
These metrics adapt to organizational infrastructures, vital for AI systems whose deployment context shapes vulnerability impact. However, AI’s evolving behavior and data interactions require more flexible metrics.

\subsubsection{Supplemental Metrics and AI}
Safety, Automatable, Recovery, and Value Density metrics add contextual insight. In AI systems, Safety is crucial as decisions affect human well-being, while Automatable is key due to large-scale automated attack risks.

\subsubsection{Notable Gaps}
While CVSS is comprehensive, several key gaps arise when applied to AI:
(1) Dynamic Nature of AI: AI systems evolve by learning from new data, altering their vulnerability landscape. Static CVSS scores may not capture this ongoing change.
(2) Indirect Impacts: AI vulnerabilities may not directly affect confidentiality, integrity, or availability but can degrade decision quality—impacts not well reflected in current CVSS metrics.
(3) Data and Model-Specific Attacks: AI-specific threats like data manipulation and model evasion lack direct analogs in traditional software, demanding new impact and exploitability metrics.

Adapting CVSS for AI thus demands systematic, dynamic assessment of AI-specific risks.
\begin{itemize}
    \item 
\textbf{Dynamic Scoring:} To address AI’s evolving nature, where model behavior changes through learning, a dynamic scoring system is needed. This involves periodic reassessment of vulnerabilities using automated tools to track behavioral shifts and update scores when significant changes occur. Key challenges include defining update frequency, maintaining management stability, and controlling the computational cost of continuous monitoring.

\item \textbf{AI-Specific Metrics:} Developing AI-specific metrics requires defining impacts and exploitabilities unique to AI systems. 
While measuring vulnerabilities through Confidentiality, Integrity, and Availability (CIA) remains important, new metrics are needed for AI software. 
These may assess \textit{susceptibility to data poisoning}, \textit{resistance to model inversion}, \textit{sensitivity to adversarial examples}, or robustness under distributional shift—factors not addressed in traditional CVSS metrics. 
The main challenge lies in quantifying such metrics consistently across diverse AI applications and industries.
\end{itemize}

This area remains open and requires further research to develop a reliable severity scoring system.

\subsection{Affected Software and Products} 
Identifies the software products, AI models, or systems vulnerable to the issue, including relevant versions, configurations, or components. Mapping a reported vulnerability to specific AI products may require further research.

\subsection{Exploitability} 
Describes how a vulnerability can be exploited or reproduced. The following factors are considered:

\begin{itemize}
    \item Technical Complexity: Level of skill and sophistication needed, e.g., crafting adversarial inputs or exploiting model design flaws.
    \item Required Privilege Level: Whether elevated privileges are needed; may include access to model parameters, training data, or inference APIs.
    \item Specific Actions Required: Steps an attacker must perform to exploit the vulnerability.
\end{itemize}

\subsection{Mitigation} 
Mitigation strategies in AI systems must account for the specific model, data, and domain. These may include software patches, model updates, or data preprocessing. In some cases, broader system-level changes are needed to improve resilience and prevent future threats.

Each mitigation technique includes an analysis of its purpose, strengths, limitations, and the specific vulnerabilities it addresses. For example, the Adversarial Example Detection technique \cite{paudice2018detection} is used to defend against adversarial machine learning attacks. Table~\ref{tab:mitigation} outlines the details of this method.

\begin{table}[ht]
    \centering
    \caption{An example of a Mitigation Technique}
    \begin{tabular}{|l|p{6cm}|}
    \hline
       \textbf{Description}  & The technique employs distance-based anomaly detection to identify adversarial examples using a subset of trusted data points. It divides these data points into two groups and trains a distance-based outlier detector for each group. In training, it establishes a pair of thresholds that are used to detect outliers. This approach helps in distinguishing between legitimate and adversarial inputs by analyzing the distances from these trusted points.\\ \hline
       \textbf{Effect} & Removes adversarial examples\\ \hline
       \textbf{Type} & Proactive (to be performed before or during training)\\ \hline
       \textbf{Tactic} & Adversarial Detection\\ \hline
       \textbf{Orientation} & Data\\ \hline
       \textbf{Target Weakness} & Inadequate Input Filtering\\ \hline
       \textbf{Target Attack} & Poisoning Attack\\ \hline
       \textbf{Pros} &  Computationally efficient for large datasets and data with a large number of features\\ \hline
       \textbf{Cons} & Not effective against label flipping poisoning attacks\\ \hline
       \textbf{Reference} & \url{https://arxiv.org/pdf/1802.03041}\\ \hline
    \end{tabular}
    \label{tab:mitigation}
\end{table}

\begin{table}[!ht]
\centering
\caption{AI Vulnerability Details}
\label{tab:vulnerability}
\scriptsize
\begin{tabular}{|>{\centering\arraybackslash}m{1.4cm}|m{6.8cm}|}
\hline
\textbf{Field} & \textbf{Details} \\
\hline
\textbf{AI-CVE ID} & 2024-1234 (example ID) \\
\hline
\textbf{Description} & The malicious actor collects data similar to the target model's training data to train multiple \textit{shadow models} that mimic the target's behavior. These models classify points as "in" (used) or "out" based on outputs. An attack model is trained on these classifications to distinguish membership, then applied to the target's outputs to infer whether specific data points were in its training set, potentially exposing sensitive information. \\
\hline
\textbf{AI System Details} & Meta data: \\
& - Name: Google Prediction API \& Amazon ML \\
& - Type: Neural Networks (CNN) \\
& - Version: 2017 \\
& - Availability: Restricted Access \\
& - Creator of the Model: Facebook (Meta) \\
& - Certification and License: Google Cloud and AWS Certification \\
& - Release Date: 05/22/2017 \\
& Architecture: \\
& - Foundation and/or Additional Model: CNN \\
& - Software and Hardware Requirements: Python3 \\
& - Dependencies: Torch7 \\
& Data: \\
& - Data Source: NIST (MNIST) \\
& - Collection: Handwritten digits collected from high school students \\
& - Quality Assessment: Normalization, Similarizing training and test set \\
& - Governance: Following 2007 Free Software Foundation, Inc. guidelines \\
& - Annotation: High school students \& United States Census Bureau \\
& Ethical: Adheres to AWS and Google Cloud Code of Conduct \\
& Environmental: Influenced by cloud machines in use \\
\hline
\textbf{Weakness Type} & Lack of appropriate privacy safeguard \\
\hline
\textbf{Root Cause(s)} & 1. Revealing entire prediction vector score\\
&2. Insufficient coarsening of the prediction vector \\
\hline
\textbf{Impact} & Expose individual data entries used in training, leading to privacy violations and the risk of leaking confidential information. \\
\hline
\textbf{AI-Severity Score} & 9.0 / 10.0 (Critical). This vulnerability exposes confidential information, has Medium Exploitability, and affects a broad range of applications. \\
\hline
\textbf{Affected Software \& Products} & Google Cloud and Amazon Web Services \\
\hline
\textbf{CPE ID(s)} & 2024/google/cloud/ModelV01, 2024/AWS/ModelV01 \\
\hline
\textbf{Exploitability} & The malicious actor needs to have (partial) access to the training data distribution and repeated query access to the target model; the model must expose full prediction vectors.\\
\hline
\textbf{Mitigation} & For this vulnerability, three mitigation strategies are advised: 

- Restricting the Prediction Vector to Top-K Classes:
Limits outputs to the top-K most probable classes, reducing information and making it harder to infer whether a data point was in the training set.

- Coarsening Prediction Precision: Lowers output granularity (e.g., via rounding or binning), obscuring how closely a data point matches the model and reducing inference accuracy.

- Differential Privacy: Adds controlled noise to data or parameters during training so outputs remain nearly unchanged with or without any individual record, protecting against membership inference and ensuring privacy.
\\
\hline
\textbf{References} & Membership Inference Attack and coarsening precision of the prediction scores\cite{7958568}, Differential Privacy\cite{leemann2024gaussian, 9745062}, Restricting Prediction to Top-k\cite{liu2023tear}. \\
\hline
\textbf{Report Date} & 03/25/2024 \\
\hline
\textbf{Reported by} & R. Shokri et al.\cite{9745062} \\
\hline
\textbf{Vendor} & Google Cloud (Alphabet), AWS (Amazon) \\
\hline
\textbf{Status} & Resolved with defensive techniques \\
\hline
\end{tabular}
\end{table}

\section{AI Vulnerability Disclosure Process} 
The AI Vulnerability Disclosure Process must reflect the unique nature and impact of AI vulnerabilities. While it can build on existing software disclosure practices, it should address AI-specific complexity through clear policies, defined roles (e.g., a security team for triage), and verification via replication or controlled testing. The process should assess impact based on data sensitivity, functionality, and exploitability, and define clear criteria for responsible disclosure.
 
Existing vulnerability management practices can be adapted to support collecting, analyzing, and defining the minimum elements. For example, establishing AI CVE Naming Authorities (CNAs)—organizations, vendors, researchers, or bug bounty providers authorized to assign CVE IDs—will help standardize AI vulnerability reporting and management.
These entities will intake vulnerability reports and assign AI-CVE IDs and determine each of the minimum elements such as \#1.AI-CVE ID, \#7. Affected Software and Products and \#8. CPE ID, \#12. References, \#13. Report Date will be determined through CNAs. 

Existing sources such as \textbf{AI/ML ATT\&CK Framework}\cite{fazelnia2022supporting, fazelnia2022attacks} can be utilizes to determine information required for the  Minimum Elements such as \#3. Weakness Type, \#4. Root Cause, \#5. Impact, and \#10. Mitigation. 
The framework categorizes AI security information into three primary components: \textit{Weaknesses}, which describe root causes, types; \textit{Attacks}, characterized by their tactics, motivation, access level, and affected models; and \textit{Mitigation}, outlining defense types, their advantages, limitations, and overall effectiveness.


\section{AI Bill of Materials to Support Minimum Elements}

AIBOM provides deep insights into the origins and composition of AI systems, enhancing component traceability and result reproducibility. It also plays a crucial role in identifying vulnerabilities and ensuring ethical standards throughout the AI product lifecycle. While resources and guidelines exist for creating SBOMs \cite{SBOM, mirakhorli2024landscape}, no standardized framework yet exists for AIBOM, revealing a critical gap in AI component management.
To our knowledge, there is no comprehensive definition of AIBOM in the literature.
In this work, we propose AIBOM, a framework that fills this gap by offering structured guidelines to document and manage elements of AI systems. AIBOM supports AIVD in systematically cataloging and assessing vulnerabilities across AI models, strengthening defenses and improving the reliability of AI deployments.

As shown in Figure \ref{fig:AIBOM}, AIBOM includes several components~\cite{AIBOM1, AIBOM2, Aibom-Squad}, described below:

\begin{figure*}[ht]
    \centering
    \includegraphics[width=\linewidth]{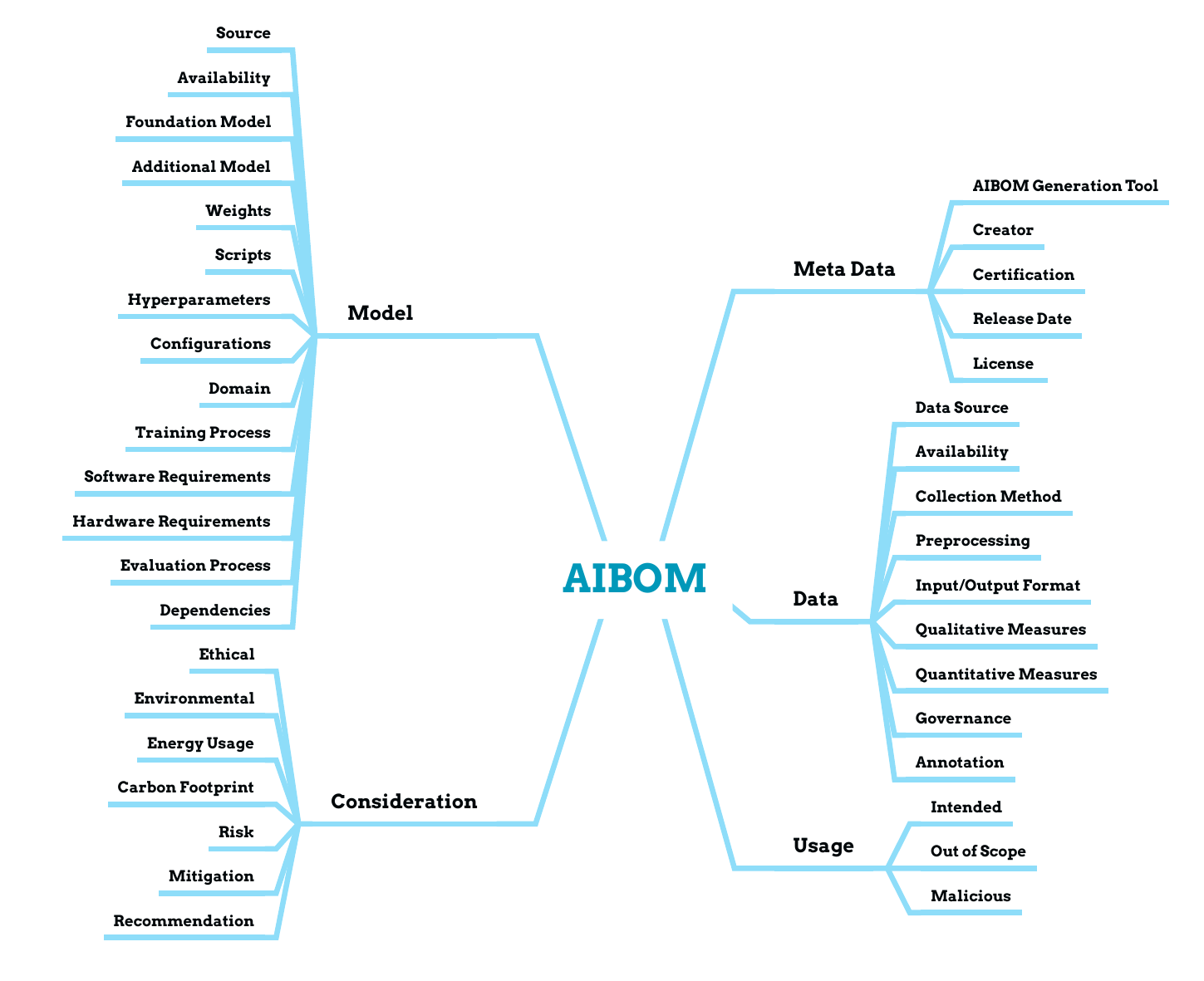}
    \caption{An overview of the proposed AIBOM}
    \label{fig:AIBOM}
\end{figure*}

\subsubsection{Meta Data}
Describes the model’s identity, origin, and authorization.
\begin{itemize}
    \item AIBOM Generation Tool: Tool used to create the AIBOM record.
    \item Creator: Individual or organization that developed the model.
    \item Certification: Official approvals or compliance credentials.
    \item Release Date: Date the model was published or deployed.
    \item License: Legal terms governing use, modification, and distribution.
\end{itemize}

\subsubsection{Model} Details of the model architecture and components include:
\begin{itemize}
    \item Source: Origin or provider of the model.
    \item Availability: Public or restricted access status.
    \item Foundation Model: Base model architecture used for adaptation.
    \item Additional Model: Supplementary models integrated with the foundation model.
    \item Weights: Pretrained or fine-tuned parameters of the model.
    \item Scripts: Code used to run the model.
    \item Hyperparameters: Key tuning parameters defining model behavior.
    \item Configurations: Settings required for execution.
    \item Domain: Application area such as healthcare, finance, or security.
    \item Training Process: Procedure followed to train the model.
    \item Software Requirements: Necessary frameworks or software versions.
    \item Hardware Requirements: Hardware resources needed for operation.
    \item Evaluation Process: Methods used to assess performance.
    \item Dependencies: Libraries and packages required by the model.
\end{itemize}

\subsubsection{Data} Details of the Data used in this model including:
\begin{itemize}
    \item Data Source: Origin of the dataset.
    \item Availability: Either public or private
    \item Collection Method: Process used for data gathering.
    \item Preprocessing: Steps applied for data preparations.
    \item Input/Output Format: Structure and format of data.
    \item Quantitative Measures: Numerical properties of data such as size.
    \item Qualitative Measures: Descriptive aspects of data such as context relevance and annotation clarity.
    \item Governance: Policies managing data access and usage.
    \item Annotation: Procedures for data labeling.
\end{itemize}

\subsubsection{Consideration} Covers ethical, environmental, and operational factors influencing the responsible use of the system.
\begin{itemize}
    \item Ethical: Compliance with ethical standards and societal values.
    \item Environmental: Broader ecological impact, e.g., data center sustainability.
    \item Energy Usage: Power consumed during training and inference.
    \item Carbon Footprint: CO$_2$ emissions from energy use.
    \item Risk: Potential ethical or technical vulnerabilities.
    \item Mitigation: Actions taken to reduce identified risks.
    \item Recommendation: Suggested improvements or best practices.
\end{itemize}

\subsubsection{Usage} Defines the intended scope and boundaries of how the model should or should not be applied.
\begin{itemize}
    \item Intended Usage: Primary application.
    \item Out of Scope: Scenarios beyond the defined use cases.
    \item Malicious Usage: Potential misuse or harmful applications of the system.
\end{itemize}

\section{Motivating Example}
We present a real-world case (tested on AWS and Google Cloud) illustrating how an AI vulnerability is identified, modeled, and managed within the AIVD framework. 
This example demonstrates AIVD’s practical use in addressing a documented vulnerability and its effectiveness in improving AI system reliability.  

\textbf{A Real-World AI Vulnerability:}
Table~\ref{tab:vulnerability} presents a real case of a membership inference attack, where a malicious actor queries an AI model to determine whether a data point was part of its training set. By exploiting prediction outputs, the attacker can infer sensitive information, violating privacy and regulatory standards. This case shows how traditional databases overlook AI-specific threats, while AIVD captures them through detailed fields like model information, severity, and mitigation, highlighting the need for a dedicated AI vulnerability database.

\section{Moving Forward}
\label{sec:moving-forward}
We must address existing gaps—for example, while “severity” can be part of AIVD, how should it be measured for AI incidents?  

Building trustworthy AI systems requires robust mechanisms for managing vulnerabilities. This section outlines the main challenges and gaps to strengthen the reliability, security, and resilience of AI deployments.

\begin{itemize}
    \item \textbf{Challenge \#1:} Complex Vulnerability Profiles: Traditional vulnerability databases like CVE and CWE are not equipped to fully capture the complexities of AI-specific issues. AI systems exhibit unique vulnerabilities due to their reliance on data quality, algorithmic complexity, and adaptive behaviors that standard vulnerability descriptors may not adequately address.
    \item \textbf{Challenge \#2:} Dynamic and Non-Deterministic Nature: The dynamic and often non-deterministic nature of AI systems complicates vulnerability identification. Traditional systems have static and predictable outputs, whereas AI outputs can change based on new data or algorithmic adjustments, making consistent vulnerability assessment challenging.
    \item \textbf{Challenge \#3:} Severity and Impact Assessment: Assessing the severity of vulnerabilities in AI systems poses significant challenges due to the variable contexts in which these systems operate. For example, the same vulnerability may have minor consequences in one scenario while causing catastrophic failures in another, depending on how the AI system is integrated and its critical role in decision-making.
    \item \textbf{Challenge \#4:} Interdependencies: AI systems often depend heavily on external data sources and dynamic learning algorithms, making them susceptible to vulnerabilities introduced through these dependencies. Identifying and tracing these vulnerabilities require a deep understanding of the interdependencies and the potential cascade effects.
    \item \textbf{Challenge \#5:} Data and Privacy Concerns: Many AI systems handle sensitive or personal data, which raises significant privacy concerns if vulnerabilities are exploited. This aspect necessitates not only tracking vulnerabilities but also ensuring compliance with data protection regulations.
    \item \textbf{Challenge \#6:} Update and Patch Management: Given the rapid development cycles of AI technologies and their applications, ensuring timely updates and patch management is a significant challenge. The continuous learning aspect of many AI systems may necessitate ongoing vulnerability assessments and updates, complicating traditional patch management strategies.
    \item \textbf{Challenge \#7:} Ethical and Bias Implications: AI systems can perpetuate or exacerbate biases present in their training data. Addressing these vulnerabilities requires not only technical solutions but also ethical considerations, which are currently not well-represented in existing vulnerability databases.
\end{itemize}

\section{Conclusions}
This paper introduces AIVD, a unified framework for managing vulnerabilities in AI systems. We define MEs for structured reporting, propose AI-CWE for AI-specific weaknesses, and present AIBOM to document AI system components. We also highlight key challenges unique to AI and illustrate how our framework enables more effective identification and mitigation of AI vulnerabilities.



\bibliographystyle{IEEEtran}
\bibliography{ref}
\textbf{Mohamad Fazelnia} is a Computer Science Ph.D. Candidate at the University of
Hawaii at Manoa at Honolulu, Hawaii, USA. His research interests include
Trustworthy Artificial Intelligence and Software Security. He received his
M.Sc. in Computer Science from University of Hawaii at Manoa. He
is a member of IEEE and the IEEE Computer Society. Contact him at
mfazel@hawaii.edu.

\textbf{Sara Moshtari} is a Postdoctoral Fellow in the Department of
Information and Computer Science at University of Hawaii at Manoa
at Honolulu, Hawaii, USA. Her research interests include Software Security,
Vulnerability Detection, and Artificial Intelligence. She received her Ph.D.
in Computer Science from Rochester Institute of Technology. Contact her at
saram23@hawaii.edu.

\textbf{Mehdi Mirakhorli} is an Associate Professor in the Department of
Information and Computer Science at University of Hawaii at Manoa at
Honolulu, Hawaii, USA. His research interests include Trustworthy Software,
Software Assurance, and Artificial Intelligence. He received his Ph.D. in
Computer Science from DePaul University. He is an associate editor of
IEEE Transactions on Software Engineering and International Journal on
Empirical Software Engineering. He is a Member of IEEE. Contact him at
mehdi23@hawaii.edu.
\end{document}